\documentclass[a4paper,fleqn,usenatbib]{mnras}

\usepackage[T1]{fontenc}
\usepackage{ae,aecompl}

\newcommand{\erf}{\operatorname{erf}}

\usepackage{graphicx}	
\usepackage{amsmath}	
\usepackage{amssymb}	
\usepackage{enumerate}  

\title[GRB duration distribution via pulse shapes]{How does the shape of gamma-ray bursts' pulses affect the duration distribution?}

\author[M. Tarnopolski]{Mariusz Tarnopolski$^{1}$\thanks{E-mail: mariusz.tarnopolski@uj.edu.pl}
\\
$^{1}$Astronomical Observatory, Jagiellonian University, Orla 171, 30-244, Krak\'ow, Poland
}

\date{Accepted XXX. Received YYY; in original form ZZZ}

\pubyear{2021}

\usepackage{newtxtext,newtxmath}

\begin{document}
\label{firstpage}
\pagerange{\pageref{firstpage}--\pageref{lastpage}}
\maketitle

\begin{abstract}
Gamma-ray bursts (GRBs) come in two types, short and long. The distribution of logarithmic durations of long GRBs is asymmetric rather than Gaussian. Such an asymmetry, when modelled with a mixture of Gaussian distributions, requires an introduction of an additional component, often associated with another class of GRBs. However, when modelled with inherently asymmetric distributions, there is no need for such a component. The cosmological dilation was already ruled out as a source of the asymmetry, hence its origin resides in the progenitors. GRB light curves (LCs) are usually well described by a series of fast-rise-exponential-decay pulses. A statistical analysis of ensembles of simulated LCs shows that the asymmetry is a natural consequence of the pulse shape and the multi-pulse character of the LCs.
\end{abstract}




\section{Introduction}
\label{sect::introduction}

Gamma-ray bursts (GRBs; \citealt{klebesadel}) are divided into two classes: {\it short} (coming from compact-object mergers) and {\it long} (gravitational collapse of massive stars). The division was put forward on the basis of a clearly bimodal duration distribution \citep{mazets,kouve93}. A third, intermediate-duration class was later proposed \citep{horvath98}. This was based on modelling the observed distribution of $T_{90}$ (i.e., time during which 90\% of the GRB's fluence is detected, starting from the moment 5\% of the fluence is accumulated) as a mixture of log-normal distributions (in practice, by modelling the distribution of $\log T_{90}$ as a mixture of Gaussian distributions). This approach was undertaken in several subsequent analyses \citep{horvath02,horvath08,zhang08,huja09,horvath10,zhang16}, usually resulting in a conclusion that three Gaussian components are needed to describe the data accurately \citep{horvath06,ripa09,horvath10,veres10,horvath18}, hence claiming the presence of three GRB classes. Some works, however, pointed at only two such components \citep{ripa12,yang16,kienlin20}.

However, the fact that three symmetric components are required to fit an apparently bimodal distribution suggests that one of the components has to be introduced to take account of the asymmetry in the data \citep{koen,tarnopolski15b}. Indeed, it was shown that a mixture of two skewed components is sufficient to satisfactorily describe the $\log T_{90}$ distributions of the BATSE\footnote{Burst And Transient Source Experiment, onboard the Compton Gamma-Ray Observatory.}, Swift, and Fermi GRBs \citep{tarnopolski16a,kwong,minaev20}. This generally holds in higher-dimensional parameter spaces as well \citep{tarnopolski19a,tarnopolski19b,tarnopolski19c}. It was sometimes noted that the distributions appear skewed, though \citep[e.g.,][]{mukherjee,toth19}, but such observation was not followed by employing asymmetric models.

While there is no need to devise a physical mechanism generating the putative intermediate GRBs (which would be bound to be a difficult task since GRBs still remain mysterious in many aspects), a question about the origin of the apparent skewness naturally arises. One possible explanation could be that the redshift distribution, when convolved with the rest-frame duration distribution, leads to asymmetry. Assuming that the rest-frame logarithmic durations are Gaussian (as was the working paradigm, and was supported by modelling the redshift-equipped GRBs; \citealt{huja09,tarnopolski16c,zhang16,zitouni18}), it was however established that redshifts can account for only a few percent of the observed skewness \citep{tarnopolski20a}. It is worth pointing out that among the $\sim 1500-3000$ GRBs observed in the BATSE, Swift, or Fermi samples each, only a total of a few hundred have an associated redshift estimate. Due to the relative smallness of the latter, the subtleties of the duration distribution can simply be not traced prominently enough \citep[cf.][]{tarnopolski19b}. An attempt to justify normality in the rest frame was made by means of the central limit theorem \citep[CLT;][]{ioka02}. The CLT, however, talks about the asymptotic case of a standardised sum of independent identically distributed (iid) random variables when the number of summands $p\to\infty$, while the Berry-Esseen theorem \citep{berry41,esseen42,zolotukhin18} states the rate of convergence for the case of a finite sum is proportional to $1/\sqrt{p}$. Additionally, sums of a small number of random variables can significantly deviate from a Gaussian distribution, especially when the variables are not iid. Note that since the observed $\log T_{90}$ are not normal, Cram\'er's theorem does not apply \citep{balazs03}. Coupled with the fact that the distribution of cosmological distances to GRBs cannot explain the skewness in the duration distribution \citep{tarnopolski20a}, its origin needs to reside at the site of emission, i.e., at the progenitors. 

The goal of this paper is to investigate whether the skewness of the duration distribution can arise due to the shape of the pulses in GRBs' light curves (LCs), and its dependence on the pulses' parameters. In Sect.~\ref{sect::methods} the LC models and simulation schemes are described. The results, presented in Sect.~\ref{sect::results}, are followed by discussion in Sect.~\ref{sect::discussion}. The conclusions are gathered in Sect.~\ref{sect::conclusions}.

\section{Models and Methods}
\label{sect::methods}

\subsection{Pulse Shape}
\label{sect::pulse}

\begin{figure}
\centering
\hspace{-0.7cm}\includegraphics[width=0.8\columnwidth]{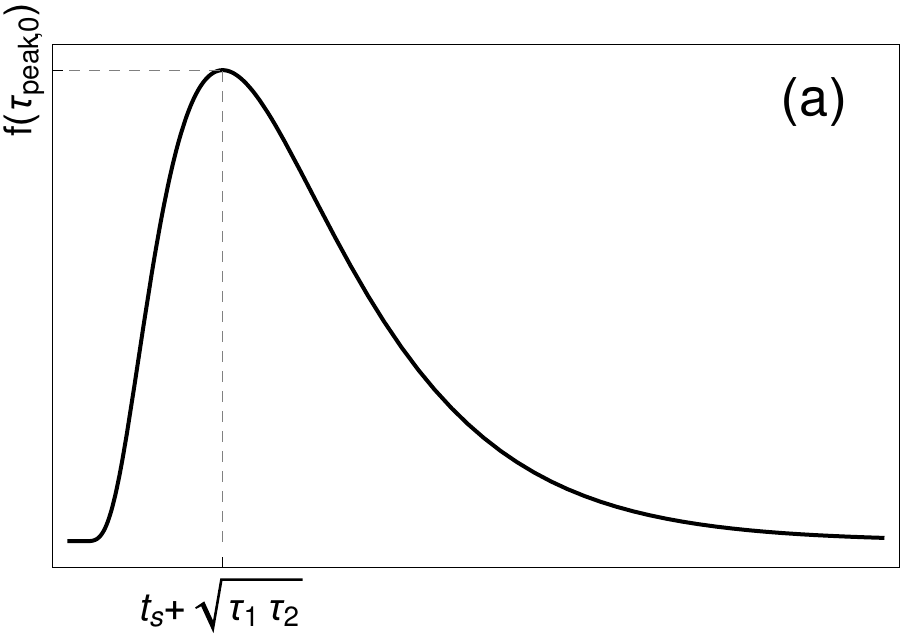}\\
\includegraphics[width=\columnwidth]{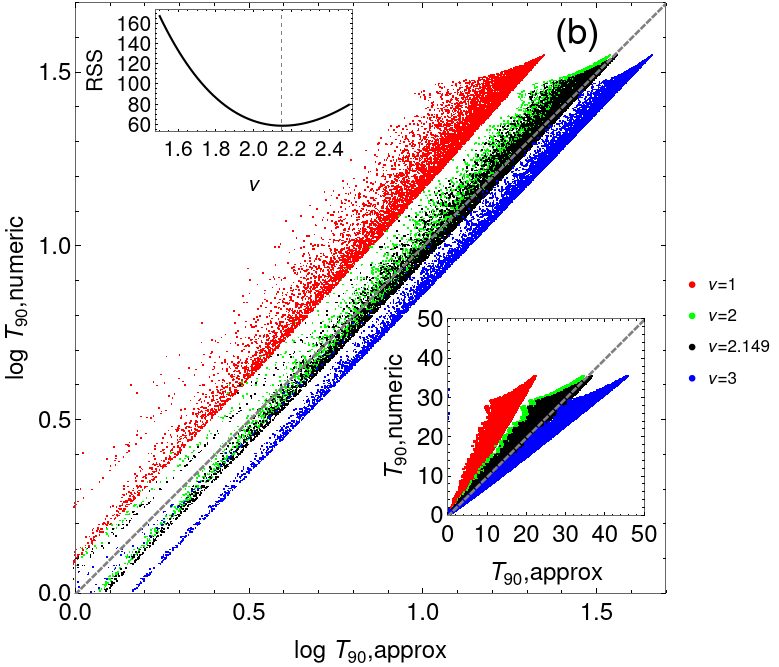}\\
\includegraphics[width=0.8\columnwidth]{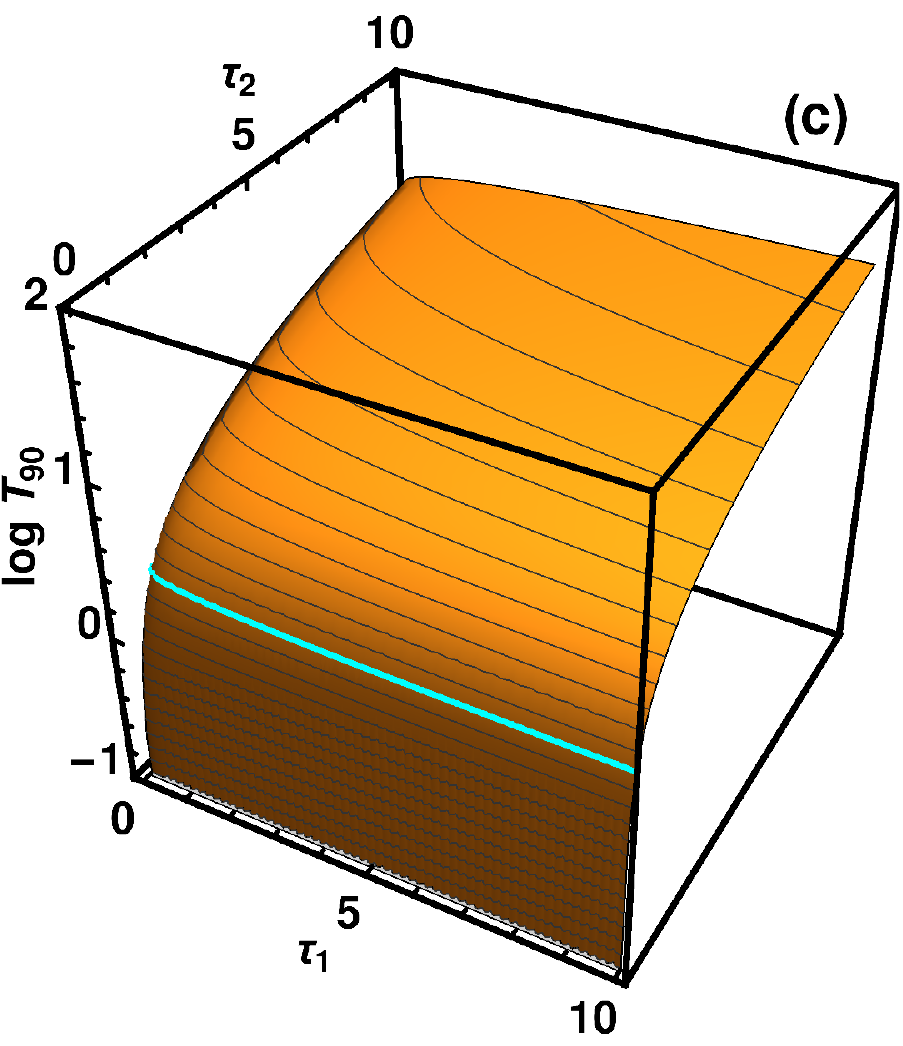}
\caption{ The FRED pulse. (a) Schematic shape. (b) Relations between the numerically computed $\log T_{90}$ and the approximation from Eq.~(\ref{eqWidth}). The lower right inset shows the same but in linear scale. The gray dashed lines denote the equality relation. Colors of the points correspond to values of $\nu$ indicated in the legend. The black points come from the value $\nu=2.149$ that minimises the RSS (upper left inset). (c) Dependence of $\log T_{90}$ on $\tau_1$ and $\tau_2$, based on Eq.~(\ref{eqWidth}) with $\nu=2.149$. The horizontal cyan line marks $T_{90}=2\,{\rm s}$. }
\label{fig2}
\end{figure}

The pulse shape, commonly referred to as the fast-rise-exponential-decay (FRED), is a smooth function \citep{norris05}:
\begin{equation}
f(t) = A\lambda\exp\left( \frac{-\tau_1}{t-t_s} - \frac{t-t_s}{\tau_2} \right)
\label{eq4}
\end{equation}
for $t\geqslant t_s$, where $t_s\in\mathbb{R}$ is the start time of the pulse, and $A,\lambda,\tau_1,\tau_2>0$. It is illustrated in Fig.~\ref{fig2}(a). Consider $t_s=0$ for simplicity for a moment. Setting $A\coloneqq\left[ 2\lambda\tau_{\rm peak,0}K_1(2\mu) \right]^{-1}$ normalises the pulse: $\int_0^{+\infty}f(t)dt=1$, where $\lambda=\exp\left( 2\mu \right)$, $\mu=\sqrt{\tau_1/\tau_2}$, $\tau_{\rm peak,0}=\sqrt{\tau_1 \tau_2}$, and $K_1$ is the modified Bessel function of the second kind \citep[cf. 3.324.1 in][]{grad07}. If the starting time $t_s\neq 0$, then $\tau_{\rm peak} = t_s + \tau_{\rm peak,0}$, and the integration yields $\int_{t_s}^{+\infty}f(t)dt=1$ (with $A$ unchanged).

To derive $T_{90}$, define the cumulative flux:
\begin{equation}
F(t) = \int\limits_{t_s}^t f(t')dt',\quad t\geqslant t_s,
\label{eq2}
\end{equation}
which, due to the normalisation, increases from zero to unity when $t\to\infty$. Then solve for $t_1$ and $t_2$ such that $F(t_1)=0.05$ and $F(t_2)=0.95$, and as per the definition: $T_{90}=t_2-t_1$.

The parameter space is $\tau_1,\tau_2>0$. Imposing the condition $T_{90}>2\,{\rm s}$ for long GRBs results in a slightly more constrained region (i.e., $\tau_2\gtrsim 1$), but produces a sharp cutoff in the distribution of $\log T_{90}$, and hence is not introduced (there is an overlap between the short and long GRB classes in the duration distributions; cf. \citealt{tarnopolski15a} and references therein).

The profile in Eq.~(\ref{eq4}) does not lead to a closed-form expression for the cumulative flux, Eq.~(\ref{eq2}), hence one can only proceed numerically, following the procedure outlined above. One can, however, utilise an approximate approach. The width $w$ of the FRED pulse, defined as the interval between the times $t$ when the pulse values obey $f(t)/f(\tau_{\rm peak,0})={\rm e}^{-\nu}$, $\nu>0$, is \citep{norris05}
\begin{equation}
w = \nu\tau_2\left( 1+\frac{4\mu}{\nu} \right)^{1/2}.
\label{eqWidth}
\end{equation}
\citet{norris05} used $\nu=1$, while \citet{hakkila14,hakkila18} set $\nu=3$ to estimate the pulse duration. To find the value of $\nu$ that leads to the best approximation of $T_{90}$, $10^4$ pulses were simulated with $\tau_i$ uniformly distributed in $(0,10)$, and then were numerically integrated to obtain the reference values of $T_{90}$. Next, the widths $w$ were computed as a function of $\nu$. Since the primary focus herein is on the distribution of $\log T_{90}$, the residual sum of squares (RSS) of the numeric and approximate values of $\log T_{90}$ was minimised. The relations between the estimated durations are displayed in Fig.~\ref{fig2}(b). The minimal RSS was obtained for $\nu=2.149$, and hence can be employed for single-pulse LCs.

The dependence of $T_{90}$ on $\tau_1$ and $\tau_2$ is depicted in Fig.~\ref{fig2}(c) using Eq.~(\ref{eqWidth}). It shows that the dependence on $\tau_1$ is weak, and $T_{90}$ is governed by the decay phase of the pulse. Indeed, expressing Eq.~(\ref{eqWidth}) as $w=\nu\sqrt{\tau_2^2+\frac{4}{\nu}\tau_1^{1/2}\tau_2^{3/2}}$, it is clear that $\tau_1$ only comes in with a leading power of $1/4$, while $w$ depends roughly linearly on the term with $\tau_2$. When $\tau_1\to 0$, $w\to \nu\tau_2$, so the pulse can be arbitrarily short.

\subsection{Waiting Time Distribution (WTD)}
\label{sect::WTD}

The distribution of waiting times, $\Delta t>0$, is a random variable modelled, following \citet{guidorzi15}, as
\begin{equation}
P(\Delta t) = (2-\alpha)\beta^{2-\alpha}\left( \beta + \Delta t \right)^{-(3-\alpha)},
\label{eq5}
\end{equation}
where $0\leqslant\alpha<2$ is responsible for the clusterisation of pulses, $\beta>0$ is a characteristic waiting time at which the power law with index $(3-\alpha)$ breaks. It is properly normalised as a probability density function, $\int_0^{+\infty}P(\Delta t)d(\Delta t)=1$. 

\subsection{Number of Pulses}
\label{sect::number}

\begin{figure}
\centering
\includegraphics[width=\columnwidth]{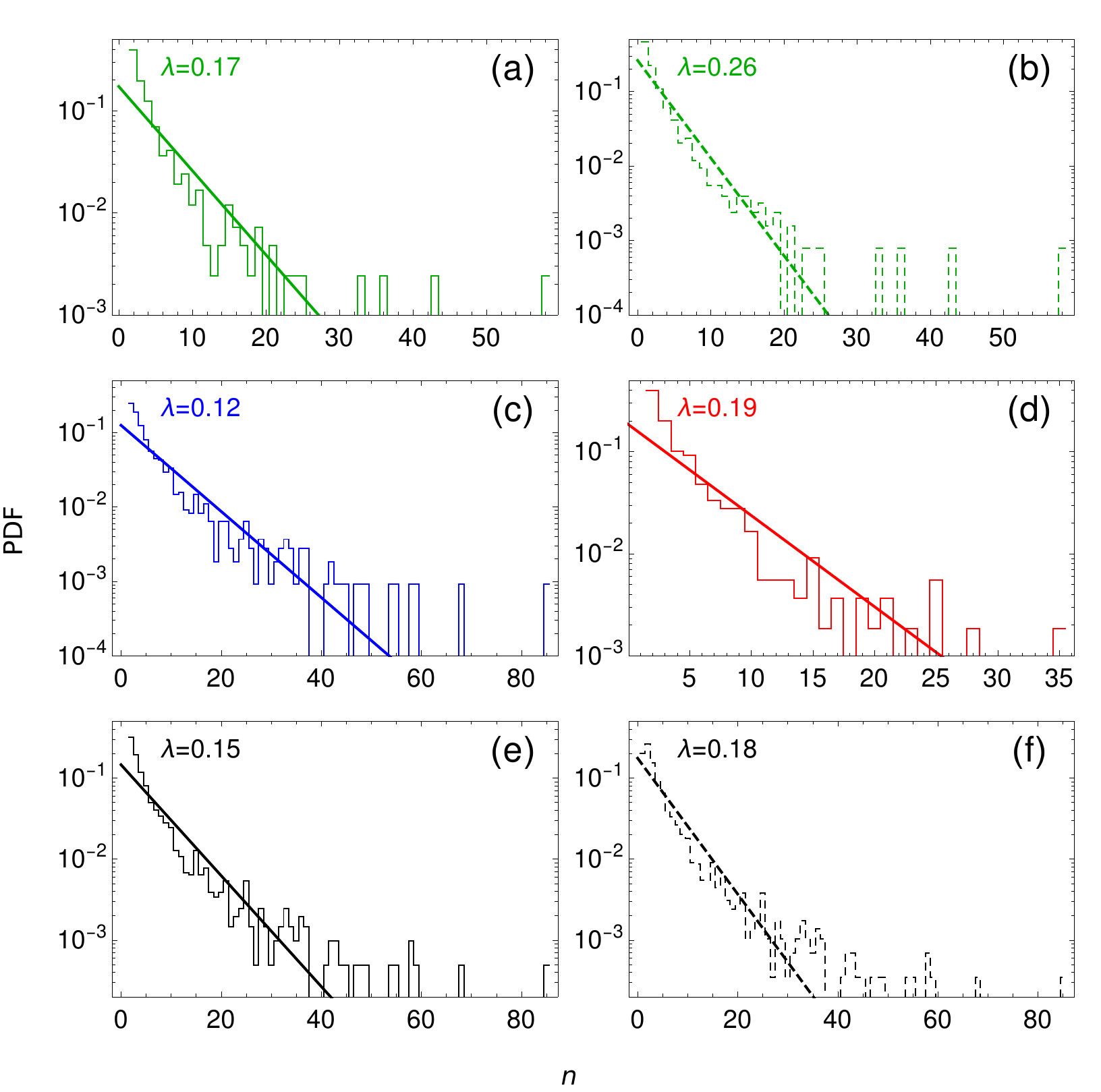}
\caption{ Distributions of the number of pulses $n$ in prompt LCs (histograms). Straight lines are the best-fit geometric distributions (maximum loglikelihood method) from Eq.~(\ref{eq7}). Panels (a), (c) and (d) show the Swift, BATSE, and Fermi samples, respectively, from \citet[][priv. comm.]{guidorzi15}. Panel (b) displays the updated (as of 30 November 2020) list of Swift GRBs. Panel (e) refers to the joint sample from (a), (c), (d), while panel (f) shows the joint sample with the updated Swift sample, i.e. data from panels (b), (c), (d). }
\label{n_pulses}
\end{figure}

\citet{guidorzi15} used a peak search algorithm \citep{mepsa} to obtain the number of pulses $n$ in BATSE (1089 GRBs), Swift (418 GRBs), and Fermi (544) samples. Their distributions are displayed in Fig.~\ref{n_pulses}(a), (c), (d) (C.~Guidorzi, priv. comm.). Additionally, the sample of Swift GRBs updated up to 30 November 2020 (1273 GRBs) is also employed [Fig.~\ref{n_pulses}(b)]. The joint samples are displayed in Fig.~\ref{n_pulses}(e) and (f).  

The logarithmic plots in Fig.~\ref{n_pulses} exhibit an approximately linear decline over a wide range of $n$, with only occasional instances of $n\gtrsim 25-40$. Therefore, the overall distribution of the number of pulses $n\in\mathbb{N}$ in an LC is a random variable that can be adequately modelled as a discrete exponential (geometric) distribution. The fits were performed via the maximum loglikelihood method \citep{kendall}. The instances of very high number of pulses in the samples can be safely neglected in the simulations, since they are rare enough to not affect the duration distribution significantly.						

The number of pulses $n\in\mathbb{N}$ is modelled in two ways:
\begin{enumerate}
\item as a discrete uniform distribution:
\begin{equation}
f_{\mathcal{U}}(n) = 
\begin{cases}
 \frac{1}{n}  & 1\leqslant n\leqslant N \\
 0 & n>N
\end{cases};
\label{eq6}
\end{equation}
\item as a geometric distribution:
\begin{equation}
f_{\mathcal{G}}(n) = 
\begin{cases}
 \left( 1-\lambda \right)^n\lambda & n\geqslant 0 \\
 0 & n<0
\end{cases},
\label{eq7}
\end{equation}
parametrised with $0\leqslant\lambda\leqslant 1$.
\end{enumerate}
The mean of the geometric distribution from Eq.~(\ref{eq7}) is $1/\lambda-1$, which for $\lambda = 0.18$ [Fig.~\ref{n_pulses}(f)] is $\approx 4.5$. The uniform distribution, Eq.~(\ref{eq6}), is employed as a testbed, to compare the resulting duration distributions.
To keep the mean of the two models the same, the maximal number of pulses in the uniform distribution is set to $N=9$.

\subsection{Skewness}
\label{sect::skew}

The skewness $\gamma$ (third standardised moment) of the distribution of long GRBs' $\log T_{90}$ is: $\gamma = -0.25$ for BATSE \citep{tarnopolski16a}, $\gamma = -0.20$ for Fermi and $\gamma = -0.63$ for Swift \citep{tarnopolski20a}\footnote{Fermi: $(\mu,\sigma,\lambda)=(1.83, 0.65, -1.45)$. Swift: $(\mu,\sigma,\lambda)=(2.25, 0.94, -2.8)$. For completeness, BATSE: $(\mu,\sigma,\lambda)=(1.89, 0.61, -1.35)$. }. These are the skewnesses of the long-GRB component of a mixture of two skew-normal ($\mathcal{SN}$) distributions,
\begin{equation}
f_{\mathcal{SN}}(x) = \frac{2}{\sigma}\varphi\left(\frac{x-\mu}{\sigma}\right)\Phi\left(\lambda\frac{x-\mu}{\sigma}\right),
\label{eqSN}
\end{equation}
where $\varphi(x) = 1/\sqrt{2\pi} \exp(-x^2/2)$ is the PDF of a standard normal distribution, and $\Phi(x) = \int_{-\infty}^x \varphi(t) dt = \frac{1}{2} \left[ 1+\erf( x/\sqrt{2}) \right]$ is its cumulative distribution function. Skewness of the $\mathcal{SN}$ distribution is $\gamma = \frac{4-\pi}{2} \left(\delta\sqrt{2/\pi}\right)^3 \left(1-2\delta^2/\pi\right)^{-3/2}$, where $\delta=\lambda/\sqrt{1+\lambda^2}$.

For all three catalogs, $\gamma<0$, meaning that the left-hand tail is heavier than the right-hand tail (i.e., there is an excess of {\it shorter} durations compared to the {\it longer} ones). The distribution of $\log T_{90}$ is best modelled with a mixture of just two skewed components in all cases. It will thence be tested if modelling an ensemble of LCs can yield the observed values of skewness of the long-GRB component.

\subsection{Skewness vs. asymmetry}

The skewness is commonly interpreted as a measure of asymmetry. Indeed, a symmetric distribution has necessarily a skewness of zero, but the converse is not always correct: a distribution with one tail short and fat, and the other long and thin can be obviously asymmetric, but their contributions to the skewness can cancel out, leading to an overall value of zero (see Appendix \ref{appA} for an example). Therefore, a notion relating directly to the property of asymmetry needs to be invoked. One possibility is the {\it distance skewness} of a random variable $X$ \citep{szekely01}, with respect to a location $\theta$:
\begin{equation}
dSkew(X) = 1-\frac{E||X-X'||}{E||X+X'-2\theta||},
\label{}
\end{equation}
whose sample version is expressed as
\begin{equation}
dSkew(X) = 1-\frac{\sum_{i,j}||x_i-x_j||}{\sum_{i,j}||x_i+x_j-2\theta||}.
\label{}
\end{equation}
For a symmetric distribution, $dSkew$ is equal to zero, and is bounded from above by unity otherwise. For an $\mathcal{SN}$ distribution from Eq.~(\ref{eqSN}) it can be formulated as
\begin{equation}
dSkew(X) = 1-\frac{\iint_{-\infty}^{\infty}|x-y|f_{\mathcal{SN}}(x)f_{\mathcal{SN}}(y)dxdy}{\iint_{-\infty}^{\infty}|x+y-2\theta|f_{\mathcal{SN}}(x)f_{\mathcal{SN}}(y)dxdy},
\label{}
\end{equation}
which can be easily computed numerically for the BATSE, Fermi, and Swift catalogs, given the best-fit parameters, resulting in 0.013, 0.018, and 0.115, respectively. The value of $\theta$ is taken as the location of the mode.

\subsection{Simulation setups}
\label{sect::simul}

The duration $T_{90}$ of the FRED pulse depends on two parameters, $\tau_1$ and $\tau_2$ [Eq.~(\ref{eq4})]. Therefore, $T_{90}$ is a random variable dependent on the distribution of these parameters. To examine the impact of the LCs' shapes on the duration distribution, the following simulations were performed.

First, single-pulse LCs were modelled:
\begin{enumerate}
\item\label{si} FRED pulse with a uniform bivariate distribution in the $(\tau_1,\tau_2)$ parameter space. The constraint $0<\tau_i<10$ is imposed;
\item\label{sii} a Gaussian distribution truncated on the positive interval, $\mathcal{TN}_{(0,\infty)}(5,2)$, independently for $\tau_1$ and $\tau_2$. 
\end{enumerate}
The threshold $T_{90}>10^{-3}\,{\rm s}$ is introduced to prevent unreasonably short durations, rarely occuring in the tails of the distributions and not observed in any GRB catalog, from significantly affecting the simulated distributions. This manual intervention is introduced to keep the simulations clean. It is a shortcoming of the FRED model, which is an empirical one that allows for arbitrarily short bursts to occur (cf. Sect.~\ref{sect::pulse}). A physically motivated model should ensure such cases do not arise \citep[it warrants a much more detailed physics-based follow-up, e.g.,][that is outside the scope of the work herein]{zhang05}. A different way to tackle this issue would be to put more stringent constraints on the parameters, but for the sake of simplicity cases with $T_{90}<10^{-3}\,{\rm s}$ are considered artifacts and simply discarded.

For multi-pulse LCs, the WTD from Eq.~(\ref{eq5}) is employed, with $\alpha=1$ and $\beta=6$ set based on the fittings of \citet{guidorzi15}. The number of pulses is drawn from the distributions in Eq.~(\ref{eq6}) and Eq.~(\ref{eq7}), in combination with the $\tau_i$ parameter distributions from setups \ref{si}--\ref{sii}. Henceforth, the following cases are considered:
\begin{enumerate}
\setcounter{enumi}{2}
\item\label{siii} setup \ref{si} and Eq.~(\ref{eq6});
\item\label{siv} setup \ref{si} and Eq.~(\ref{eq7});
\item\label{sv} setup \ref{sii} and Eq.~(\ref{eq6});
\item\label{svi} setup \ref{sii} and Eq.~(\ref{eq7}).
\end{enumerate}
Additionally, to assess the impact of the WTD on the skewness, a continuous uniform distribution, $\Delta t\sim \mathcal{U}\left( 0,50 \right)$, is tested as well:
\begin{enumerate}
\setcounter{enumi}{6}
\item\label{svii} setup \ref{si}, $\Delta t\sim \mathcal{U}\left( 0,50 \right)$, and Eq.~(\ref{eq6});
\item\label{sviii} setup \ref{sii}, $\Delta t\sim \mathcal{U}\left( 0,50 \right)$, and Eq.~(\ref{eq7}).
\end{enumerate}

For each setup in points \ref{si}--\ref{sviii}, 2500 LCs were generared (reflecting the number of GRBs in current catalogs), durations $T_{90}$ were computed, and the skewness and $dSkew$ of the resulting distribution of $\log T_{90}$ was recorded. This was repeated 100 times, eventually giving a distribution of skewness in each case \ref{si}--\ref{sviii}. The simulation scheme for multi-pulse LCs is explained in Appendix~\ref{appB}, and each pulse needs to be assigned a weight, $\omega_i$. The weights are drawn from a continuous uniform distribution, $\mathcal{U}(0,1)$. The LC is then normalised to unity for convenience.


\subsection{Multi-pulse bursts}

For multi-pulse FRED bursts, Eq.~(\ref{eqWidth}) cannot be directly used as the pulses might overlap, so a straightforward sum of the widths $w$ will not give the duration. The total duration could be, in principle, decomposed as the rise time from the first ${\rm e}^{-\nu}$ intensity to the peak of the first pulse, the sum of the waiting times, and the decay time from the peak of the last pulse to the last ${\rm e}^{-\nu}$ intensity. However, the portions of the decay and rise times will depend for each LC on the particular realisation of the number of pulses $n$ and the waiting times $\Delta t$, hence an approximate treatment similar to that in Sect.~\ref{sect::pulse} is unfeasible. Therefore, a precise computation of $T_{90}$ requires a numerical treatment. For the sake of uniformity, single-pulse bursts are also analysed numerically like the multi-pulse ones.

\section{Results}
\label{sect::results}

\subsection{Single-pulse bursts}

\begin{figure*}
\centering
\includegraphics[width=\textwidth]{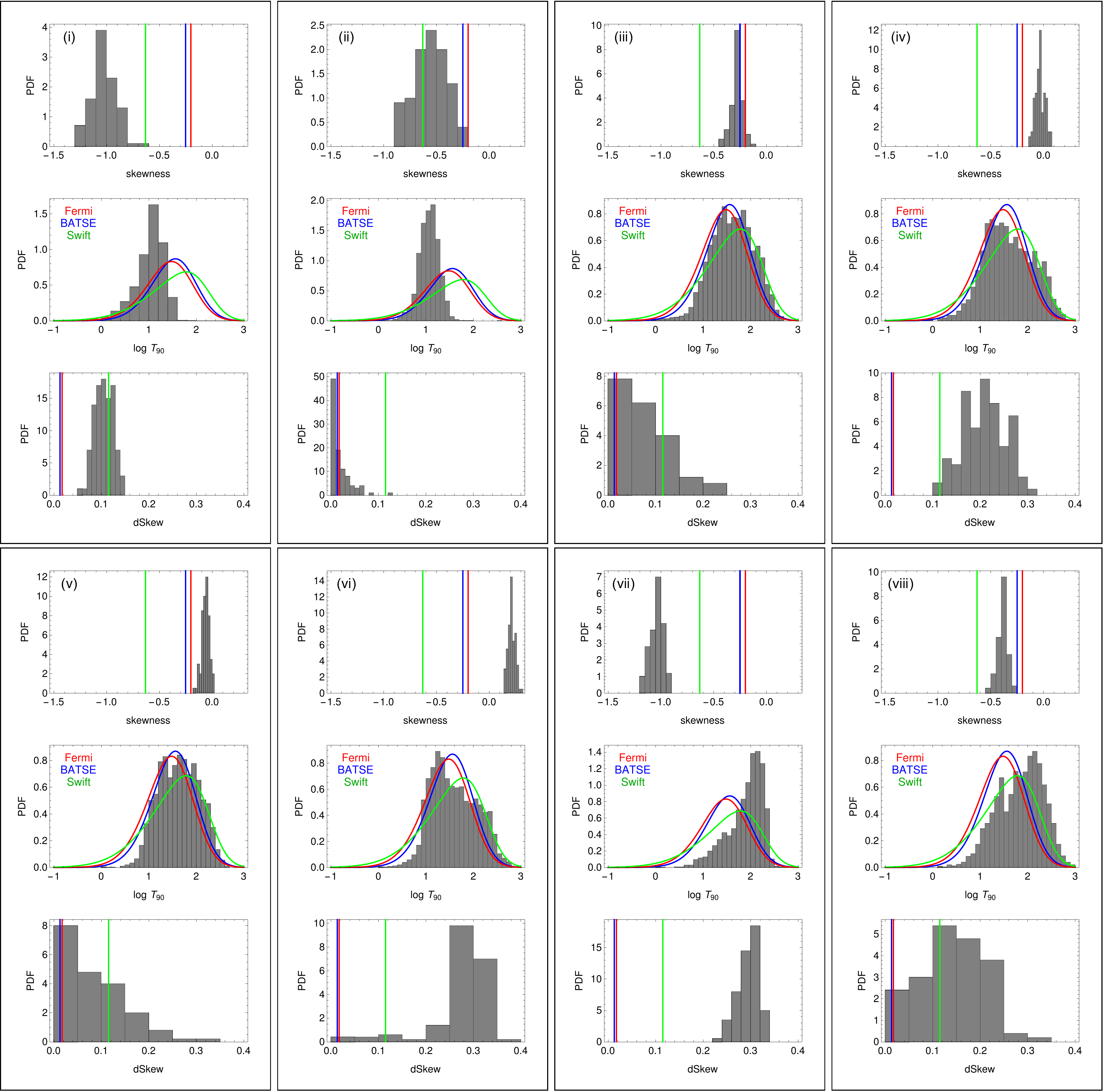}
\caption{ Distributions of skewness and distance skewness. Labels correspond to setups in points \ref{si}--\ref{sviii} from Sect.~\ref{sect::simul}, in particular: \ref{si} and \ref{sii} -- single-pulse bursts, \ref{siii}--\ref{sviii} -- multi-pulse bursts. In each framed panel, the upper plot shows the distribution of skewness, with values for the BATSE, Fermi and Swift samples marked for reference with blue, red and green vertical lines, respectively. The middle plots show typical duration distributions obtained in each setup (more precisely, the distributions yielding skewness values at the mode of the skewness distribution). The curved lines represent the $\mathcal{SN}$ distributions fitted to the BATSE, Fermi and Swift GRB samples \citep{tarnopolski16a,tarnopolski20a} in corresponding colours. The lower plots display the distribution of distance skewness. }
\label{fig3}
\end{figure*}

Single-pulse bursts cannot model appropriately the LCs of all actual GRBs, though they are a useful testbed for uncovering the properties of these building blocks. Therefore, in Fig.~\ref{fig3} are displayed the distributions of skewness and $dSkew$ for setups \ref{si} and \ref{sii}, with values corresponding to the Fermi, BATSE, and Swift samples indicated, as well as exemplary $\log T_{90}$ distributions, overlaid with best fits to the actual data from the three catalogs. Overall, the bursts are too short to match the data, which is natural due to their duration limited by the employed range of parameters. However, in both setups the resulting skewness is confidently negative, while $dSkew$ indicate asymmetric distributions. In setup \ref{si}, the uniform distributions of $\tau_i$ -- the only parameters governing the pulse shape -- gave skewed and asymmetric duration distributions, implying that the shape of the pulse itself gives rise to negative skewness. Its magnitude is slightly diminished when a more concentrated distributions of $\tau_i$ are considered in setup \ref{sii}, hence showing hints of a robust effect.

\subsection{Multi-pulse bursts}

Of particular interest are setups \ref{siv} and \ref{svi}, i.e., those in which the WTD is modelled after \citet{guidorzi15} with a physically motivated form, with the distribution of the number of pulses $n$ inferred as a geometric distribution based on the samples of Swift, BATSE, and Fermi GRBs [Eq.~(\ref{eq7})]. Setup \ref{siv}, with uniform distributions of $\tau_i$, yields distribution of skewness concentrated around zero. However, the $dSkew$ values are significantly positive, implying asymmetry, which is clearly visible in the simulated $\log T_{90}$ distributions [Fig.~\ref{fig3}]. This motivates the use of $dSkew$ as the measure of asymmetry instead of skewness. Setup \ref{svi}, in turn, similarly as in the single-pulse cases, shifts the skewness to higher values, slightly increasing the asymmetry as well.

Setups \ref{siii} and \ref{sv}, invoking a uniform distribution of the number of pulses [Eq.~(\ref{eq6})] for comparison, exhibited similar behaviour of the skewness distribution. The $dSkew$ values are more concentrated near zero, consistent with the values obtained for the three GRB catalogs. Finally, setup \ref{svii}, with all distributions used for modelling being uniform, gives the most skewed and asymmetric duration distributions, as was the case of single-pulse bursts in setup \ref{si} as well. Setup \ref{sviii}, in turn, results in skewness and $dSkew$ values broadly consistent with all values from data, and the overall shapes of $\log T_{90}$ distributions resemble those from the catalogs.

\section{Discussion}
\label{sect::discussion}

\subsection{Summary}

Simulations of the LCs as composed of FRED pulses, conducted herein, did not invoke explicitly any GRB subclasses: the parameters $\tau_i$ were drawn from single populations given by setups \ref{si} or \ref{sii}, the WTD was either uniform [in setups \ref{svii} and \ref{sviii}], or asymptotically power law [Eq.~(\ref{eq5})], and the number of pulses $n$ was geometric [Eq.~(\ref{eq6})] or uniform [Eq.~(\ref{eq7})]. Therefore, none of these settings, in conjunction with the pulse shape [Eq.~(\ref{eq4})] could trivially lead to skewed, asymmeric or bimodal duration distribution. Yet such were the outcomes for most of the examined setups. It was also highlighted that an asymmetric distribution can exhibit zero skewness, hence treating the latter as a measure of asymmetry can lead to incorrect conclusions.

By the employed simulation scheme it was demonstrated that the observed properties of the $\log T_{90}$ distribution (skewness, asymmetry, sometimes bimodality) arise naturally as a result of the properties of the building blocks of the LCs. The rise of a FRED pulse is most likely triggered by some transient process \citep{kocevski03}, e.g., injection of energetic particles into the newly forming relativistic jet, initial merging of internal shocks, or decrease of the optical depth immediately after the gravitational collapse and fallback of the ejected stellar envelope. Hence the source of the skewness observed in the Fermi, BATSE, and Swift catalogs lies in the pulse's shape and the WTD, which are directly governed by the physical processes ruling the progenitors. The distribution of the number of pulses $n$ obviously influences the shape of the duration distribution (at least its location on the duration axis). The pulses are thought to occur when relativistic shells in the jets collide. The number of shells' collisions is connected with the amount of matter fueling the jet and its spatial distribution, i.e., reflect the nature of the central engine. 

\subsection{Cosmological and instrumental effects}

For single-pulse distant GRBs only the brightest portions of LCs are observable on Earth \citep{kocevski13}. This is due to a signal-to-noise ratio increasing with redshift, which leads to obscuring larger portions of the LC for greater redshifts. Therefore, the more sensitive the detector is, the longer are the GRBs. Such an effect can dominate over the cosmological dilation, which itself was showed to account for only a few percent of the observed skewness \citep{tarnopolski20a}. Moreover, since the low-energy region of the GRB spectrum becomes undetectable when the redshift increases, the same burst observed at different redshifts would have both a different duration and spectrum. However, for multi-pulse GRBs this might be different, since the quiescent phase between the pulses would be predominantly subject to the cosmological dilation leading to longer durations. Only weak evidence for such phenomena have been reported \citep{zhang13,littlejohns14,turpin16}, hence it appears not to be decisive in the context discussed herein. 

Detector threshold clearly impacts the GRB durations that it can detect \citep{osborneII}, which in turn are positively correlated with the isotropic energy and luminosity \citep{shahmoradi13,shahmoradi15}, with a correlation coefficient $\sim 0.5-0.7$ \citep{osborneI}. This is roughly in line with the fact that brighter GRBs tend to have longer durations, as mentioned above, and are located at greater redshifts \citep{grupe13}. 

The selection effects will eventually play a role in shaping the distribution of $\log T_{90}$, however herein only the general impact of the LCs' shapes on the durations was examined to determine whether, given an ideal observing scenario, the observed skewness could be in principal attributed to the properties of the progenitors. It turns out it can, hence a physics-based research, e.g., examining if the distribution of the envelope masses of the progenitors is the source of the skewed duration distribution \citep{zitouni15}, coupled with appropriate selection and instrumental effects, ought to be conducted to give a clear and final conclusion on the topic.

\section{Conclusions}
\label{sect::conclusions}

It was established that skewness and asymmetry of the GRB duration distribution can be reached by the LC shapes altogether, without explicitly invoking any number of subclasses. Therefore, the obtained results disfavour the existence of GRB classes other than short and long. Although some models of the pulses, similar to the one employed here, were incorporated into some theoretical framework of the emission \citep{zhang05}, a thorough explanation of the LCs' shape -- their building blocks in particular -- is still desired, as well as identifying the physical origin of skewness.

\section*{Acknowledgements}

The author is grateful to Cristiano Guidorzi for sharing data on the number of pulses in GRB prompt LCs. Support by the Polish National Science Center through OPUS grant No. 2017/25/B/ST9/01208 is acknowledged.

\section*{Data Availability}
The data underlying this article will be shared on reasonable request to the corresponding author.

\bibliographystyle{mnras}
\bibliography{bibliography}



\appendix

\section{Asymmetric distribution with a zero skewness}
\label{appA}

\begin{figure}
\centering
\includegraphics[width=\columnwidth]{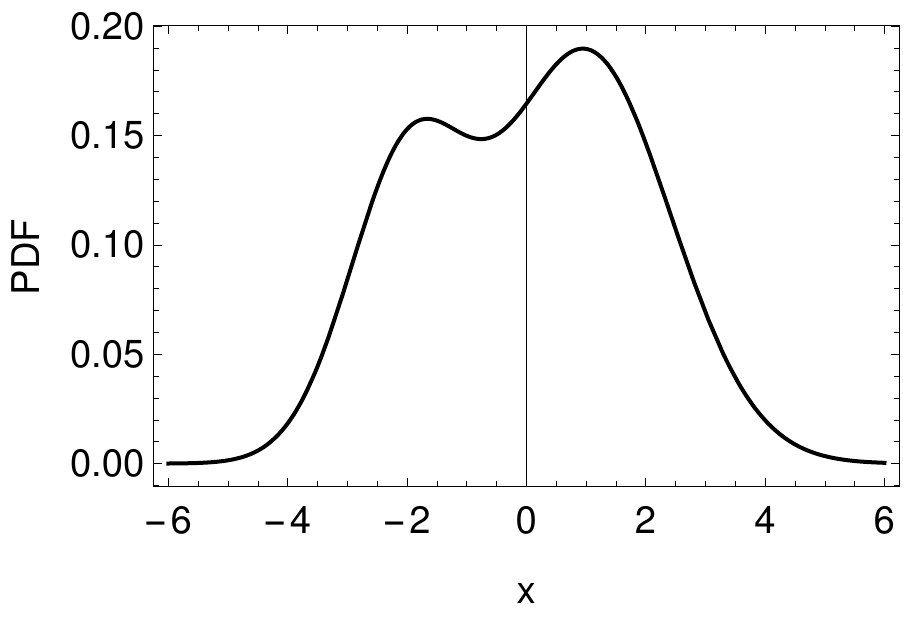}
\caption{ Example of a distribution which is asymmetric yet yields zero skewness. }
\label{figA1}
\end{figure}

A mixture of two Gaussian distributions, $\mathcal{N}(-2,1)$ and $\mathcal{N}(1,\sqrt{2})$, with weights $1/3$ and $2/3$, respectively, has a zero mean and zero skewness, yet is clearly asymmetric (Fig.~\ref{figA1}).

\section{Modelling the multi-pulse FRED bursts}
\label{appB}

\begin{figure*}
\centering
\includegraphics[width=\textwidth]{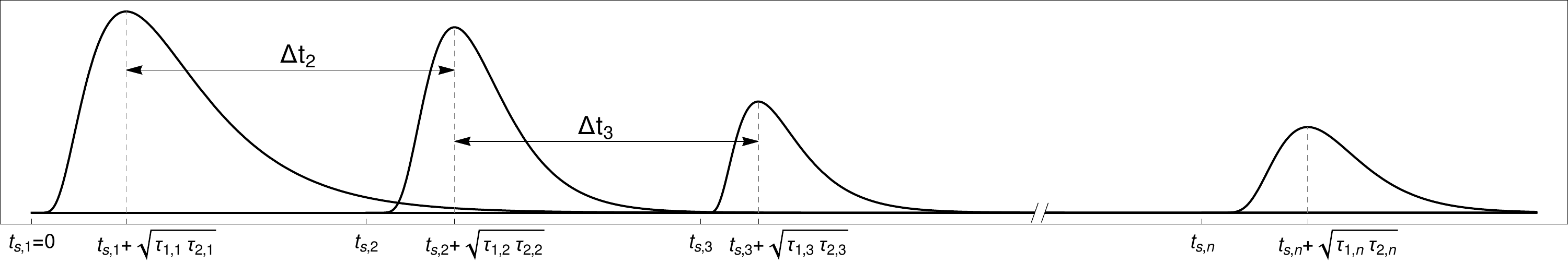}
\caption{ Parameters governing the multi-pulse FRED bursts. }
\label{figB1}
\end{figure*}

The FRED pulses' shapes and locations are described by the parameters $t_s,\tau_1,\tau_2$. To directly utilise the notion of a waiting time $\Delta t$ as the time between adjacent peaks \citep{guidorzi15}, one needs to express it in terms of these parameters. From Fig.~\ref{figB1} it follows that in general
\begin{align}
\begin{split}
\Delta t_i &= \tau_{{\rm peak},i} - \tau_{{\rm peak},i-1} \\ &= \left( \sqrt{\tau_{1,i}\tau_{2,i}}+t_{s,i} \right) - \left( \sqrt{\tau_{1,i-1}\tau_{2,i-1}}+t_{s,i-1} \right) \\ &\equiv \Delta\tau_{{\rm peak},i} + t_{s,i} - t_{s,i-1},
\end{split}
\label{eqB1}
\end{align}
where $\Delta\tau_{{\rm peak},i} = \sqrt{\tau_{1,i}\tau_{2,i}} - \sqrt{\tau_{1,i-1}\tau_{2,i-1}}$, and $t_{s,1}=0$. Therefore, the waiting times $\Delta t_i$ and the parameters $\tau_{1,i}$, $\tau_{2,i}$ ($i=2,\ldots,n$) were drawn randomly, and the starting times $t_{s,i}$ were constructed using the recursive formula
\begin{equation}
t_{s,i} = t_{s,i-1} + \Delta t_i - \Delta\tau_{{\rm peak},i}.
\label{eqA2}
\end{equation}

\bsp	
\label{lastpage}
\end{document}